\documentclass[aoas,preprint]{imsart}

\RequirePackage[OT1]{fontenc}
\RequirePackage{amsthm,amsmath}
\RequirePackage[numbers]{natbib}
\RequirePackage[colorlinks,citecolor=blue,urlcolor=blue]{hyperref}
\usepackage{graphicx}
\usepackage{subfigure}
\usepackage{mathrsfs}
\usepackage{amsthm}
\usepackage{algorithm}
\usepackage{algorithmic}
\usepackage{booktabs}
\usepackage{multirow}
\usepackage{color}

\arxiv{arXiv:0000.0000}

\startlocaldefs
\numberwithin{equation}{section}
\theoremstyle{plain}
\newtheorem{thm}{Theorem}[section]
\newtheorem{remark}{Remark}[section]
\newtheorem{myAssumption}{Assumption}[section]
\endlocaldefs

\begin{document}

\begin{frontmatter}
\title{Statistical Simulator for the Engine Knock}
\runtitle{Statistical Simulator for the Engine Knock}

\begin{aug}
\author{\fnms{Xun} \snm{Shen}\thanksref{m1,m2}\ead[label=e1]{shenxun@ism.ac.jp}},
\author{\fnms{Tinghui} \snm{Ouyang}\thanksref{m3}\ead[label=e2]{ouyang.tinghui@aist.go.jp}}
\author{\fnms{Jiancang} \snm{Zhuang}\thanksref{m1,m2}
\ead[label=e3]{zhuangjc@ism.ac.jp}
\ead[label=u1,url]{http://bemlar.ism.ac.jp/zhuang/}}
\and
\author{\fnms{Chanyut} \snm{Khajorntraidet}\thanksref{m4}\ead[label=e4]{chanyut.k@eat.kmutnb.ac.th}}

\runauthor{X. Shen et al.}

\affiliation{The Graduate University for Advanced Studies\thanksmark{m1}, The Institute of Statistical Mathematics\thanksmark{m2}, National Institute of Advanced Industrial Science and Technology\thanksmark{m3} and King Mongkut's University of Technology North Bangkok\thanksmark{m4}}

\address{Address of the First and Third authors\\
10-3 Midori-cho, Tachikawa, Tokyo 190-8562, Japan\\
\printead{e1}\\
\phantom{E-mail:\ }\printead*{e3}\\
\printead{u1}}

\address{Address of the Second author\\
2-3-26 Aomi, Koto-ku, Tokyo, 135-0064, Japan\\
\printead{e2}}

\address{Address of the Fourth author\\
1518 Wong Sawang, Bang Sue, Bangkok, 10800 Thailand\\
\printead{e4}}

\end{aug}

\begin{abstract}

This paper proposes a statistical simulator for the engine knock based on the Mixture Density Network (MDN) and the accept-reject method. The proposed simulator can generate the random knock intensity signal corresponding to the input signal. The generated knock intensity has a consistent probability distribution with the real engine. Firstly, the statistical analysis is conducted with the experimental data. From the analysis results, some important assumptions on the statistical properties of the knock intensity is made. Regarding the knock intensity as a random variable on the discrete time index, it is independent and identically distributed if the input of the engine is identical. The probability distribution of the knock intensity under identical input can be approximated by the Gaussian Mixture Model(GMM). The parameter of the GMM is a function of the input. Based on these assumptions, two sub-problems for establishing the statistical simulator are formulated: One is to approximate the function from input to the parameters of the knock intensity distribution with an absolutely continuous function; The other one is to design a random number generator that outputs the random data consistent with the given distribution. The MDN is applied to approximate the probability density of the knock intensity and the accept-reject algorithm is used for the random number generator design. The proposed method is evaluated in experimental data-based validation.   

\end{abstract}



\end{frontmatter}

\section{Introduction} 
Even though the electrification of vehicle powertrain has picked up an aggressive speed in the automotive industry, the majority will still be Hybrid Electric Vehicles (HEVs) at least for the next half century\cite{Lewis}. In the HEVs, the power is provided by two sources: a motor and an internal combustion engine\cite{Chau}. Thus, to realize the economic and green transportation system of the future society, it is of great importance to increase efficiency and reduce pollutant emissions of the engine by the combustion control\cite{Eriksson}. A big issue in the combustion control is how to make a trade off between efficiency and abnormal combustion such as the engine knock\cite{Bares}. The combustion process is dramatically influenced by the Fuel Injection Timing (FIT) in the diesel engines or the Spark Advance Timing (SAT) in the gasoline engines\cite{Payton_CST}. Generally, the FIT or SAT with the Maximal Brake Torque (MBT) is chosen to realize the highest efficiency of the energy transformation\cite{Kiencke}. However, the FIT or SAT with the MBT might cause a self-ignition in an unburned mixture, called end gas. The self-ignition often brings the pressure oscillations in the cylinder chamber of the engine. The phenomenon of the pressure oscillation is coupled with a metallic sound from the wall of the cylinder chamber. Thus, it is named the knock event\cite{Eriksson}, or knock for abbreviation.  Small amount of slight knock is good for the combustion efficiency while frequent knock can cause serious cylinder damages and a increase of pollutant emission\cite{Kiencke}. The trade off is to operate the engine at a boundary or borderline with a specific knock probability\cite{Eriksson}. To realize the boundary control or borderline control, many researches have converged to a statistical feedback control framework\cite{Payton_CST}. The statistical feedback controller does not response to the raw measurements directly. Firstly, the statistical properties of the measurements such as the probability density function or probability mass function are estimated. Then, the feedback controller adjusts the control input to realized the desired probability distribution. To evaluate the performance of the statistical knock controller, it needs a batch of experiments since one good case does not mean that the controller would succeed statistically\cite{Eriksson}. The repeats of the experiment tests causes a big expense.  To decrease the cost on the experimental tests, it is useful to develop a knock simulator to test the statistical knock controller instead of the experimental tests. 

The state-of-the-art methodologies of designing the knock simulator can be classified into 2 main streams. The first stream is the combustion physical model-based simulation\cite{Di}. In the combustion physical model-based simulator, the determinacy of the combustion is described by the Wiebe function and Livengood-Wu integration. The stochastic part is modeled by adding some noises on the deterministic part. Then, the cylinder pressure obtained from the 'stochastic' heat release profile exhibits cycle-to-cycle variations. The knock signal is obtained by setting a threshold for the peak cylinder pressure or the integration of the cylinder pressure. If the threshold is surpassed, the cycle is identified as a knock cycle, otherwise, it is a cycle without knock. The other stream of the knock simulation methodology is recognized as stochastic process-based or Markov process-based simulation\cite{Payton}. The stochastic process-based simulator focuses particularly on the knock signal itself. It investigates the statistical properties of the knock intensity, which can be calculated from the pressure oscillation\cite{Panzani}, or measured by vibration sensor\cite{Boubai}. The probability distribution of the knock intensity varies as the input of the engine or the operating condition of the engine changes. The common method is to calibrate the map from the input and the operating condition to the parameters of the probability distribution. Then, statistical simulation is done according to the parameters of the probability distribution\cite{Payton}. The map adopts the linear model or exponential model. The probability distribution of the knock intensity adopts Exponentially modified Gaussian distribution\cite{Spelina1,Spelina2}. However, the exist method still have drawbacks:
\begin{itemize}
    \item The conventional simulators only approximate a conditional mean and variance of the data assuming that the data is single Gaussian. Under the single Gaussian assumption, only a very limited statistics can be represented by the conventional simulators. If the mixture distribution model is adopted, it is able to obtain a more complete description of the probability distribution;
    \item The relation from input and operating condition to the parameters of the knock intensity distribution is not well addressed with a simple model, polynomial models or exponential models. The error of the model would bring bias between the distribution of the actual data and the simulated data.
    \item Except for the establishment of the knock simulator, the question that how to validate the simulator quantitatively has not been answered yet in the exist researches.
\end{itemize}

To address above drawbacks, the knowledge of statistics is necessary. This paper proposes a statistical knock simulator based on the Mixture Density Network (MDN) and the accept-reject method. At first, the statistical analysis is conducted with the experimental data. Regarding the knock intensity as a stochastic process, it is statistical independent and identically distributed under identical input. Moreover, the distribution, or the parameters of the distribution, can be regarded as a function of the input. The MDN is applied to approximate the function from input signal to the knock intensity distribution. The accept-reject algorithm is used to generate knock intensity according to the MDN-based knock intensity distribution model. The proposed method is evaluated in experimental data-based validation.

The rest of the paper is organized as followings: Section \ref{section:problem} presents the data analysis results, assumptions about the knock intensity and the problem formulation. Then, section \ref{section:proposed} describes the proposed statistical simulator. Section \ref{section:validation} gives the results of experimental data-based validations. Finally, section \ref{section:conclusion} concludes the whole paper.

\section{Experimental Data Analysis and Problem Formulation}
\label{section:problem}
\subsection{Experimental Details and Data Pre-processing}
\begin{table}
\centering
\caption{Specification of the Diesel engine 4JK1-TC.}
\begin{tabular}{c c}
\hline
\hline
Engine system & Detail \\
\hline
No. of cylinders & 4-cylinder \\
Arrangement & In line \\
Displacement & 2499 cc \\
Compression ratio & 18:1 \\
Maximum output & 87kW \\
Maximum torque & 280 Nm\\
Idle speed & 800 rpm\\
Maximum engine speed  & 4200 rpm\\
Fuel system & Direct injection\\
Ignition system & Compressed ignition \\
\hline
\hline
\end{tabular}
\label{4JK1-TC}
\end{table}

The data used in this study was collected from the 4JK1-TC, a common-rail direct injection inter-cooled turbo-diesel engine. The specifications of the 4JK1-TC is listed in Table. \ref{4JK1-TC}. During the collection of the database of the engine tests, the 4JK1-TC was operated at Wide Open Throttle (WOT) under a variety of speed, manifold pressure, and FIT conditions. These variables affect an engine's propensity to knock, and their values were chosen to span both a range of knock intensities as well as the normal operating envelope of the engine. The database essentially defines a multi-dimensional grid in which engine speed is varied in increments of 400 rpm in the range 800-2000 rpm, and a variety of manifold pressures are applied, typically within the range 3 bar to 8 bar. Besides, the air/fuel ratio is constantly fixed at 14.6. At each point in the grid, the FIT corresponding to Borderline knock (BL) was identified, and the FIT was then varied typically within the range BL-4 to BL+2 degrees.  

In total, a series of 118 different operating points were considered. Three data sets were recorded at each point, containing data for 300 consecutive combustion events in each individual record. Each record has the high-speed in-cycle recordings of cylinder pressure for the cylinder $\# 1$. The knock intensity metric can be obtained by calculating the power spectral density of the cylinder pressure data over the resonant frequency from the in-cylinder pressure. The details of the calculation process can be referred to \cite{Shen2019}. 

\subsection{Statistical Analysis of Data Sets}

\subsubsection{Statistical independence}
\begin{figure}[!htbp]
\centering
\includegraphics[width=4.5in]{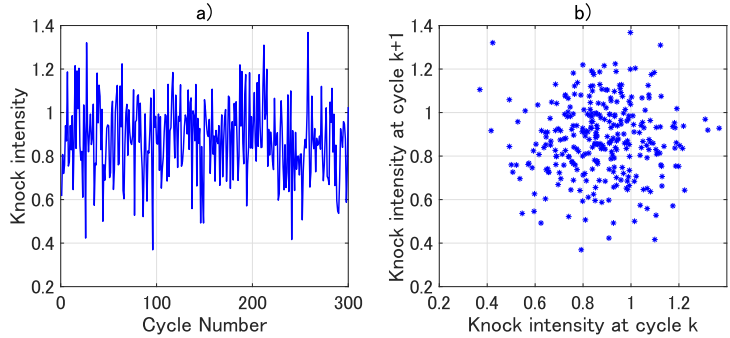}
\caption{Time series example of the knock intensity: a) Cycle-to-cycle variation in the knock intensity; b) Scatter plot of the knock intensity.}
\label{fig:knock_cycle_scatter}
\end{figure}
Fig. a) shows the cycle-to-cycle knock intensity series obtained from the 300 cycles taken from cylinder $\# 1$ of the engine operated at 800 rpm, wide open throttle, and one degrees of fuel injection timing relative to the borderline knock condition. The figure shows that the knock intensity has random cycle-to-cycle variation even under steady operating condition. The scatter plot of Fig. b) gives additional evidence to that the knock intensity has little or no prior cycle correlation. A more quantitative analysis is obtained by calculating the autocorrelation function $r(k)$ of the knock intensity $KI_{i}$ for different cycle lags $k$ as
\begin{equation}
\label{eq_r_k}
r(k)=\frac{\sum_{i=1}^{n-k}(KI_{i}-\mathbf{E}\{KI\})(KI_{i+k}-\mathbf{E}\{KI\})}{\sum_{i=1}^{n}(KI_{i}-\mathbf{E}\{KI\})^2}.
\end{equation}

\begin{figure}[!htbp]
\centering
\includegraphics[width=4.5in]{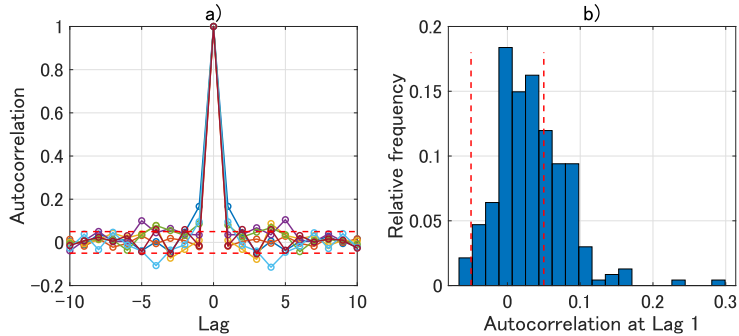}
\caption{Autocorrelation function: a) Five examples; b) Lag 1 autocorrelation function all data sets.}
\label{fig:auto_corr}
\end{figure}

The results are summarized in Fig. \ref{fig:auto_corr}. As shown in Fig. \ref{fig:auto_corr} a), most of the correlations for cycle lags $k\geq1$ fall within the $95\%$ probability interval which means there is not statistical significance in most cases. However, there are very rare cases in which the correlation falls outside these limits. Although this means that the prior cycle correlation is statistically significant, it should also be noticed that the absolute magnitude of this correlation is still very small, less than 0.2 in all cases. Fig. \ref{fig:auto_corr} b) shows the results of autocorrelation function at Lag 1 for the entire database. In most cases, the autocorrelation function at Lag 1 is below 0.2. The rare cases with value larger than 0.2 are all the operating conditions with the increased instability of the combustion process. According to the above statistical independence analysis results and discussions, the assumption about the knock intensity under a given operating condition can be summarized as:
\begin{myAssumption}
\label{Ind_assum1}
The knock intensity is independently and identically distributed random variables for a given operating condition.
\end{myAssumption}

\subsubsection{Probability distributions}

\begin{figure}[!htbp]
\centering
\includegraphics[width=4.5in]{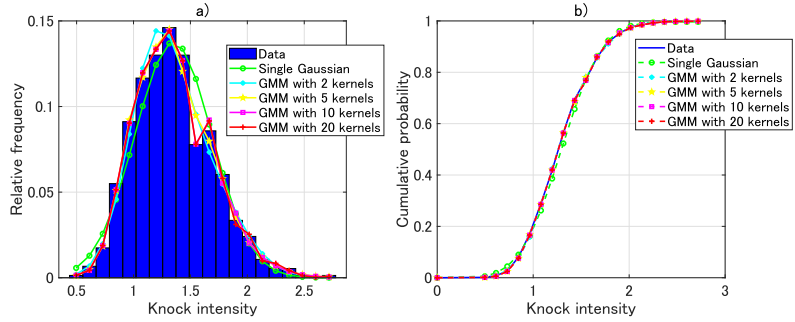}
\caption{The comparison of the distribution fit performance in one operating condition (engine speed: 800 rpm, manifold pressure: 6 bar, relative fuel injection timing: +1 degree.): a) Relative frequency; b) Cumulative probability}
\label{fig:com_dis}
\end{figure}
\begin{figure}[!htbp]
\centering
\includegraphics[width=4.5in]{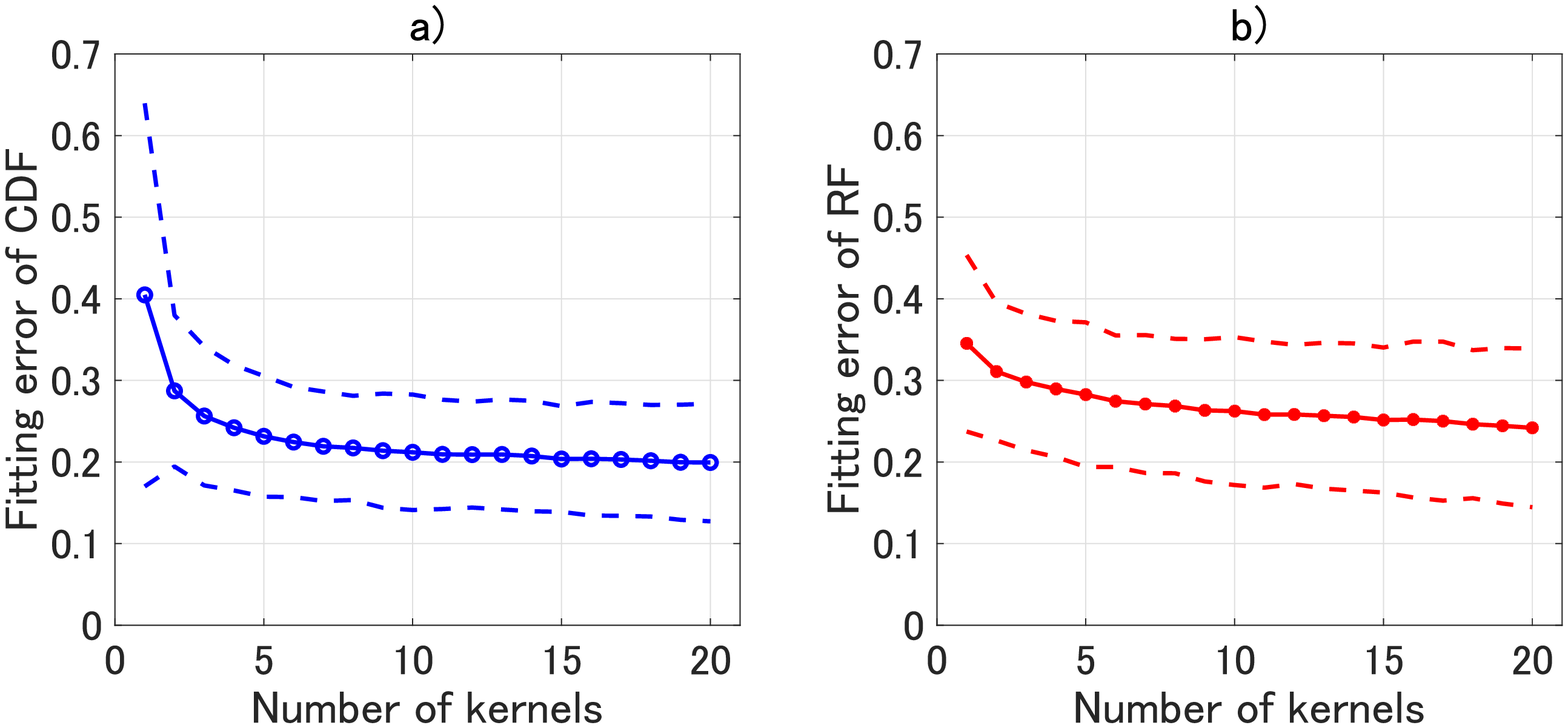}
\caption{The comparison of the fitting error: a) Cumulative Probability (CP) the blue solid line with circles represents the mean error of all database and the dashed lines represent the bounds of 99$\%$ confidence interval; b) Relative Frequency (RF) the red solid line with points represents the mean error of all database and the dashed lines represent the bounds of 99$\%$ confidence interval.}
\label{fig:com_fit}
\end{figure}

The results of the distribution fit performance are summarized in Fig. . The comparisons of the relative frequency and cumulative density function of the knock intensity are shown which exhibits that Gaussian mixture models have better performance on fitting the relative frequency and cumulative probability. Besides, a more concise measure of the goodness of fitting is to compare the fitting errors of each model calculated from historical data for all operating conditions. The fitting error is defined as:
\begin{equation}
\label{eq:fitting_error}
E=\sqrt{\sum_{i=1}^{N_o}(o_i-\hat{o}_i)^2}
\end{equation}
where $o_i$ denotes the $i$-th measured output variable in a data record of length $N_o$, $\hat{o}$ represents the model-estimated output\cite{Devore}. Small $E$ values indicate good model fits. Here, defining $o_i=F(z_i)$ as the cumulative probability or relative frequency, and $\hat{o}_i=\hat{F}(z_i)$, where $z_i$ denotes the measured value for the knock intensity. The approach was replicated here, giving the results shown in Fig. \ref{fig:com_fit}. As the number of kernels increases, both fitting errors decrease, sharply from one to two and then gently from three. Thus, for the trade-off of accuracy and complexity, the number of kernels can be chosen as values from 2-5.  
According to the above statistical analysis of the knock intensity distribution, the following assumption can be summarized:
\begin{myAssumption}
\label{Ind_assum2}
Gaussian mixture models have better accuracy on fitting the distribution of the knock intensity than single Gaussian model. As numbers of the used kernels increases, the fitting accuracy increase. 
\end{myAssumption}

\begin{figure}[!htbp]
\centering
\includegraphics[width=4.5in]{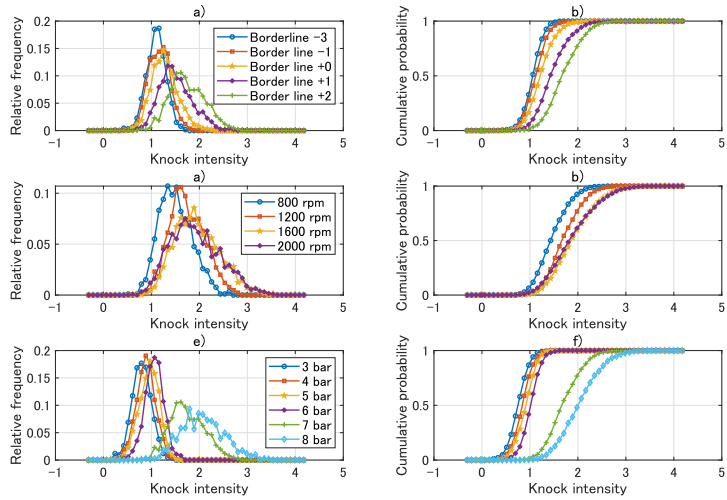}
\caption{The characteristics of knock intensity's distribution: a) relative frequency (1200 rpm, 7 bar); b) cumulative probability (1200 rpm, 7 bar); c) relative frequency (7 bar, Border line +2); d) cumulative probability (7 bar, Border line +2); e) relative frequency (1200 rpm, Border line +2); f) cumulative probability (1200 rpm, Border line +2).}
\label{fig:inp_out}
\end{figure}

\begin{figure}[!htbp]
\centering
\includegraphics[width=4.5in]{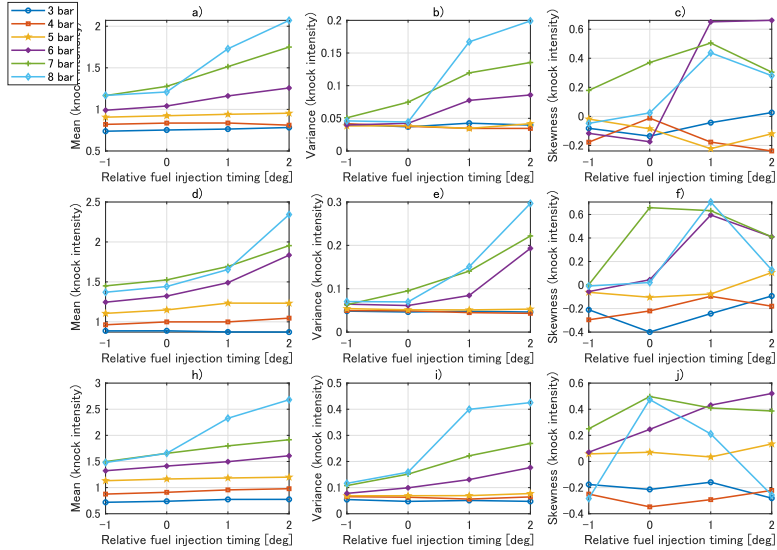}
\caption{The characteristic parameters of knock intensity's distribution (mean, variance, and skewness): a-c) 1200 rpm; d-f) 1600 rpm; g-i) 2000 rpm.}
\label{fig:inp_out_all}
\end{figure}

Fig. \ref{fig:inp_out} a) shows a example of the knock intensity relative frequency curves for a range of different fuel injection timing taken from cylinder $\#$1 of the engine operated at 1200 rpm and manifold pressure of 7 bar. The corresponding cumulative probability curves are plotted in Fig. \ref{fig:inp_out} b). As the fuel injection timing is advanced, the curves of relative frequency and cumulative probability shifts to the right side which means the means and modes of the distribution clearly increase. This reflects the strong influence of fuel injection timing on the engine's propensity to knock intensity. Similar results are obtained at a range of different engine speed from cylinder $\#$1 of the engine operate at manifold pressure of 7 bar and borderline +2 or at a range of different manifold pressure from cylinder $\#$1 of the engine operated at 1200 rpm and borderline +2. These observations highlight the fact that the fuel injection timing, manifold pressure and engine speed all influence the knock intensity distribution strongly. More details are shown in Fig. \ref{fig:inp_out_all}. The characteristic parameters of knock intensity's distribution, such as mean, variance and manifold pressure, vary as the fuel injection timing, manifold pressure and engine speed change. According to the above discussion, the assumption on the relationship between the operating condition and the knock intensity distribution can be summarized:
\begin{myAssumption}
\label{Ind_assum3}
The parameters of the knock intensity distribution are affected by the input and operating condition of engine, such as fuel injection timing, manifold pressure and engine speed. The parameter vector can be regarded as a function of the input and operating condition of engine.
\end{myAssumption}

\subsection{Problem Formulation}
Denote the parameter vector of the engine operating conditions as $u\in\mathcal{R}^d$ and the knock intensity as $y\in\mathcal{R}$. According to Assumption \ref{Ind_assum1}, \ref{Ind_assum2} and \ref{Ind_assum3}, probability density of $y$ depends on $\theta$ and expressed as $p(y|u)$. The addressed problem is to obtain an approximation probability density $\hat{p}(y|u)$ of the probability density $p(y|u)$ using a training sample set $T_N=\{(u_1,y_1),...,(u_i,y_i),...,(u_N,y_N)\}$. The problem is expressed as
\begin{equation}
\label{eq:p_a}
\hat{p}^*(y|u) = \arg\max_{\hat{p}} \sum_{i=1}^{N}\log \hat{p}(y_i|u_i).
\end{equation}

\section{Proposed Statistical Knock Simulator}
\label{section:proposed}
\begin{figure}[!htbp]
\centering
\includegraphics[width=4.5in]{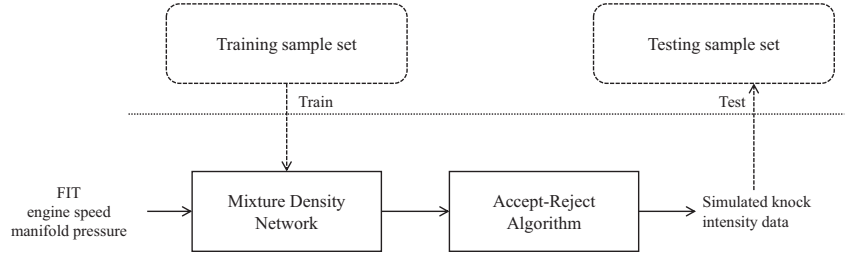}
\caption{Schematic of the proposed statistical knock simulator.}
\label{fig:Overview}
\end{figure}

The schematic of the proposed simulator is illustrated in Fig. \ref{fig:Overview}. Firstly, a Mixture Density Network is trained using the training sample set. The obtained Mixture Density Network outputs the parameter vector of the approximated probability density function of the knock intensity. The parameter vector is conditioned on the input and operating conditions of the engine such as FIT, engine speed and manifold pressure. Then, an Accept-Reject algorithm is applied to generate the knock intensity data corresponding to the conditioned probability density function. The generated data will be checked whether their distribution are consistent with the data in testing sample set. 

\subsection{Mixture Density Networks}
In the framework of MDN, a mixture model is used to approximate the probability density of the target data instead of using the Gaussian distribution model. The mixture model has the flexibility to model completely general distribution functions \cite{McLachlan}. The approximate probability density of the target data is then represented as a linear combination of kernel functions in the form
\begin{equation}
\label{eq:kernel}
\hat{p}(y|u)=\sum_{i=1}^{m}w_i(u)\phi_{i}(y|u)
\end{equation}
where $m$ is the number of components in the mixture model. The parameters $w_i(u)$ are called mixing coefficients which is a prior conditional probabilities (conditioned on $u$) of the knock intensity $y$ having been generated from the $i$-th component of the mixture. Note that the mixing coefficients $w_i(u)$ are also taken to be functions of the input vector $u$ and $\forall i\in\{1,...,m\},\forall u\in\mathcal{R}^d$ satisfies the following constraints
\begin{equation}
\label{eq:w_constraint1}
w_i(u)>0,
\end{equation}
\begin{equation}
\label{eq:w_constraint2}
\sum_{i=1}^{m}w_i(u)=1.
\end{equation}
The functions $\phi_{i}(y|u)$ denote the conditional density of the knock intensity $y$ for the $i$-th kernel. Various choices for the kernel functions are possible. In this research, the chosen kernel functions are Gaussian of the form
\begin{equation}
\label{eq:gau_ker}
\phi_{i}(y|u)=\frac{1}{2\pi\delta_i(u)}\exp\{-\frac{\|y-\mu_i(u)\|^2}{2\delta_i^2(u)}\}
\end{equation}
where $\mu_i(u)$ and $\delta_i(u)>0,\forall u\in\mathcal{R}^d$ represent the centre and variance of the $i$-th kernel respectively. In principle, a Gaussian mixture model with kernels given by (\ref{eq:gau_ker}), can approximate any given density function to arbitrary accuracy if the mixing coefficients and the Gaussian parameters (means and variances) are correctly chosen\cite{McLachlan}. Therefore, the representation given by (\ref{eq:kernel}) and (\ref{eq:gau_ker}) is completely general. Formally, as given in \cite{Botev}, the above discussions can be summarized in Theorem \ref{GMM}.
\begin{thm}
\label{GMM}
For a given $u$, $\forall i\in{1,...,m}$, let $w_i(u)=\frac{1}{m}$ and let $\delta_{i}(u)=\delta(u)$ for be such that $\lim_{m\rightarrow\infty}\delta(u)=0$ and $\lim_{m,N\rightarrow\infty}m\sqrt{\delta(u)}=\infty$. Assume that $p''(y|u)$is a continuous square-integrable function. The integrated squared bias and integrated variance of the Gaussian mixture density model expressed by (\ref{eq:kernel}) and (\ref{eq:gau_ker}) have asymptotic behavior 
\begin{equation}
\label{eq:mean_bias}
\int\|\mathrm{E}\{\hat{p}(y|u)\}-p(y|u)\|^2dy=\frac{1}{4}\delta^2(u)\|p''(y|u)\|^2+o(\delta^2(u)), \ \ \ \ \ \ m,N\rightarrow\infty
\end{equation}
and
\begin{equation}
\label{eq:variance_bias}
\int\text{Var}_f\{\hat{p}(y|u)\}dy=\frac{1}{2m\sqrt{\pi\delta(u)}}+o(\frac{1}{m\sqrt{\delta(u)}}), \ \ \ \ \ m, N\rightarrow\infty.
\end{equation}
respectively. The first-order asymptotic Approximation of Mean Integrated Squared Error(AMISE), is thus given by 
\begin{equation}
\label{eq:AMISE}
\text{AMISE}\{\hat{p}(y|u)\}(\delta(u))=\frac{1}{4}\delta^2(u)\|p''(y|u)\|^2+\frac{1}{2m\sqrt{\pi\delta(u)}}.
\end{equation}
The asymptotically optimal value of $\delta(u)$ is minimizer of the AMISE
\begin{equation}
\label{eq:opt_delta}
\delta^*(u)=(\frac{1}{2m\sqrt{\pi}\|p''(y|u)\|^2})^{2/5},
\end{equation}
giving the minimum value
\begin{equation}
\label{eq:min_AMISE}
\text{AMISE}\{\hat{p}(y|u)\}(\delta^*(u))=N^{-4/5}\frac{5\|p''(y|u)\|^{2/5}}{4^{7/5}\pi^{2/5}}.
\end{equation}
\end{thm}
The simple proof of Theorem \ref{GMM} is given in \cite{Wand}.
\begin{remark}
\label{remark_GMM}
Theorem \ref{GMM} gives the upper boundary of AMISE when using the Gaussian mixture density model expressed by (\ref{eq:kernel}) and (\ref{eq:gau_ker}) since $w_i(u),\delta_i(u)$ can be optimized to get smaller AMISE instead of fixing them as two constants. Namely, if $w_i(u),\delta_i(u),\mu_i(u)$ are well estimated, the AMISE of optimal $\hat{p}^*(y|u)$ satisfies
\begin{equation}
\label{eq:min_AMISE_opt}
\text{AMISE}\{\hat{p}^*(y|u)\}\leq\text{AMISE}\{\hat{p}(y|u)\}(\delta^*(u))=N^{-4/5}\frac{5\|p''(y|u)\|^{2/5}}{4^{7/5}\pi^{2/5}}.
\end{equation}

Even the upper boundary converges to 0 if $m$ and $N$ increase to $\infty$, thus the Gaussian mixture density model $\hat{p}(y|u)$ is able to converge to the actual probability density $p(y|u)$ theoretically.
\end{remark}

On the other hand, for any given $u$, the mixture model expressed by (\ref{eq:kernel}) gives a general formalism for modelling an arbitrary conditional density function $p(y|u)$. Denote the parameter vector of the mixture model as
\begin{equation}
\label{eq:para_theta}
\theta(u)=[w_1(u),...,w_m(u),\mu_1(u),...,\mu_m(u),\delta_1(u),...,\delta_m(u)]^T.
\end{equation}
The parameter vector $\theta(u)\in\mathcal{R}^{3m}$ is taken to be a continuous function of $u$. The continuous function can be approximated by using a conventional neural network which takes $u$ as input and $\theta$ as output. According to \cite{Huang}, conventional neural network can approximate any given continuous function to arbitrary accuracy. Formally, the approximate property is summarized in Theorem \ref{cnn}.
\begin{thm}
\label{cnn}
Given a bounded nonlinear activation function $g: \mathcal{R}\rightarrow\mathcal{R}^{3m}$ for which there exists $\lim_{\|x\|\rightarrow\infty}g(x)$, then for any $N_s$ arbitrary distinct samples $\{(u_j,\theta_j(u_j))|u_j\in\mathcal{R}^d,\theta_j(u_j)\in\mathcal{R}^{3m},j=1,...,N_s\}$, there exist $\alpha_j\in\mathcal{R}^d$, $b_j\in\mathcal{R}$, and $\beta_j\in\mathcal{R}$, $j=1,...,N_s$, such that
\begin{equation}
\label{eq:cnn_approx}
\sum_{j=1}^{N_s}\beta_jg(\alpha_j\cdot u_j+b_j)=\theta_j(u_j).
\end{equation}
\end{thm}
The proof of Theorem \ref{cnn} is given in \cite{Huang}. 

\begin{figure}[!htbp]
\centering
\includegraphics[width=4.5in]{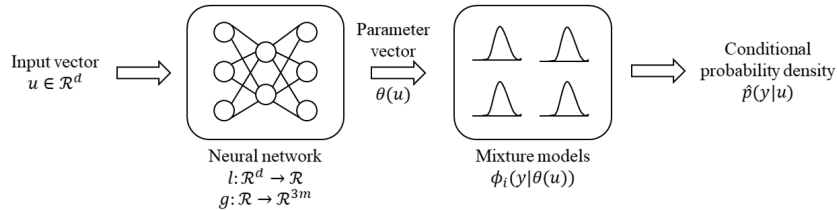}
\caption{The Mixture Density Networks consists of a feed-forward neural network and a mixture density model. The outputs of the feed-forward neural network determine the parameters in the mixture density model. The mixture model the represents the conditional probability density function of the knock intensity, eventually conditioned on the input vector of the neural network.}
\label{fig:MDN_str}
\end{figure}

Theorem \ref{GMM} and \ref{cnn} are the theoretic basis of MDN. The basic structure is illustrated in Fig. \ref{fig:MDN_str}. By choosing a mixture model with a sufficient number of kernel functions, and a neural network with a sufficient number of kernel functions, the MDN can approximate any conditional density $p(y|u)$ as closely as desired. Define the parameter of the neural network as $\theta_{NN}=[\beta_1,...,\beta_{N_s},\alpha_1,...,\alpha_{N_s},b_1,...,b_{N_s}]^T$, problem given by (\ref{eq:p_a}) can be then equivalently replaced by the problem defined as
\begin{equation}
\label{eq:prob_equi}
\theta_{NN}^*=\arg\max\sum_{i=1}^{N}\log \hat{p}(y_i|\theta(u_i,\theta(u_i,\theta_{NN}))).
\end{equation}
Therefore, the probability density $p(y|u)$ can be approximated by a Mixture Density Network. The parameter vector of the Mixture Density Network is obtained through solving problem expressed by (\ref{eq:prob_equi}) with a training sample set $T_N$. The gradient decent algorithm can be used for training the parameter vector $\theta_{NN}$\cite{McLachlan}. The obtained Mixture Density Network has a property that it has the maximal likelihood to generate the data with the identical distribution of the training sample set. 

\subsection{Accept-Reject algorithm for statistical simulation}

It is difficult to directly generate data $y~\hat{p}(y|u)$ since the approximate probability density $\hat{p}(y|u)$ is too complex. Accept-Reject method can be applied to simulate data of arbitrary probability density. The philosophy of Accept-Reject is summarized as:
\begin{thm}
\label{AR}
Let $y~\hat{p}(Y|u)$ and let $g(Y|u)$ be a simple density function (for example, uniform distribution) that satisfies $\hat{p}(Y|u)\leq Mg(Y|u)$ for some constant $M\geq 1$. Then, to simulate $y~\hat{p}(Y|u)$, it is sufficient to generate $y_g~g(Y|u)$ and $u_f~U_f|y_g=y~\mathcal{U}(0,Mg(y|u))$, until $0<u_f<\hat{p}(Y|u)$. $U_f|y_g=y$ means uniform distribution conditioned on $y_g=y$.
\end{thm}
The proof of Theorem \ref{AR} can refer to Chapter 2 of \cite{Robert}. Based on Theorem \ref{AR}, the Accept-Reject algorithm is formed which is summarized in Algorithm \ref{AR_alg}.
\begin{algorithm}[h]
\label{AR_alg}
\raggedright
\caption{Accept-Reject algorithm} 
1:\ Step 1: Generate $y_g~g(Y|u)$, $u_y~\mathcal{U}_{[0,1]}$;\\
2:\ Step 2: Accept $y=y_g$ if $u_y\leq\frac{\hat{p}(y_g|u)}{Mg(y_g|u)}$, otherwise, reject $y_g$;\\
3:\ Step 3: Return to 1 and repeat.
\end{algorithm}

\begin{remark}
\label{remark:whole}
If $\hat{p}(y|u)$ well approximates the probability density $p(y|u)$, $\forall u$, Algorithm \ref{AR_alg} can generate the data $\hat{y}$ which has the same conditional probability density of $y~p(y|u)$.
\end{remark}

\section{Validation Result}
\label{section:validation}

This section gives the validation results using experimental data set. Both steady and transient cases are concerned.

\subsection{Validation methodology}
In the validation, the data for testing is different from the data used for training the parameters of the MDN models. 

For the steady case, the data set mentioned in section \ref{section:problem} is used. The data set is firstly separated into 118 different subsets according to the operating conditions. Each subsets is given a label $i\in\{1,2,3,...,116,117,118\}$. Then, 118 different validation simulations were implemented. In each simulation, the subset $i$ was chosen as the test set and the rest subsets were training set. The MDN-based simulators with kernel numbers from 1 to 20 were validated in each simulation. Each simulator generated 500 groups of data set in each test. In every group, the number of simulated data is 900.

For the transient case, the training set is the whole data set mentioned in section \ref{section:problem}. The data of the test set was obtained in another experiment under a transient case.

\subsection{Steady case}
\begin{figure}[!htbp]
\centering
\includegraphics[width=4.5in]{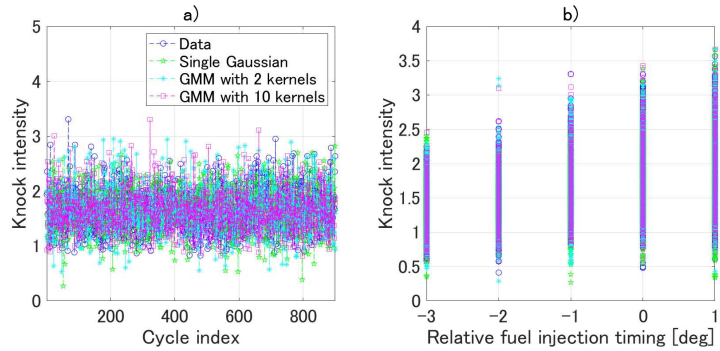}
\caption{Data of test set vs simulated data (steady case): a) Cycle index; b) Respect to relative fuel injection timing.}
\label{fig:time_series}
\end{figure}
\begin{figure}[!htbp]
\centering
\includegraphics[width=4.5in]{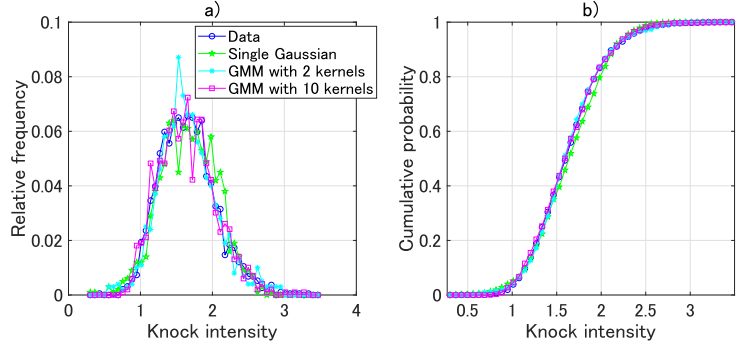}
\caption{The characteristics of data distribution: test set vs simulated data (steady case). a) Relative frequency; b) Cumulative probability.}
\label{fig:distribution_exa}
\end{figure}
\begin{figure}[!htbp]
\centering
\includegraphics[width=4.5in]{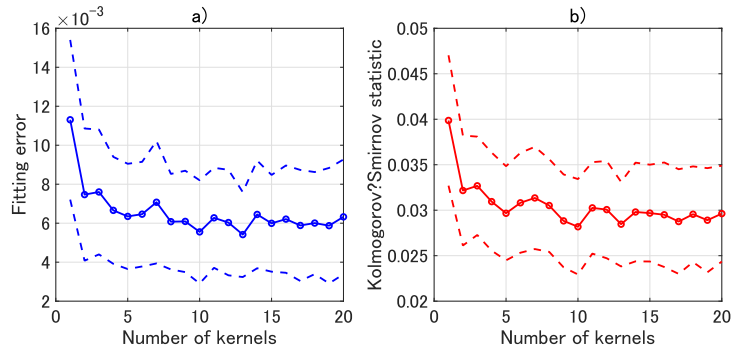}
\caption{The comparison of the simulation performance: a) Fitting error of the cumulative probability the blue solid line with circles represents the mean error of all test database and the dashed lines represent the bounds of 99$\%$ confidence interval; b) Kolmogorov-Smirnov statistic of the cumulative probability the red solid line with circles represents the mean error of all test database and the dashed lines represent the bounds of 99$\%$ confidence interval.}
\label{fig:error_steady}
\end{figure}

Fig. \ref{fig:time_series} shows the comparison between the data from the test set and the data generated by the simulators. Fig. \ref{fig:time_series} a) gives the cycle-to-cycle plots of real data, simulated data using MDN with single Gaussian, 2 Gaussian kernels, 10 Gaussian kernels respectively. The engine was operated at speed of 2000 rpm. Besides, the manifold pressure was 7 bar and the relative fuel injection timing was operated at the borderline knock condition. Fig. \ref{fig:time_series} shows results of more relative fuel injection timings, from -3 to 1. Intuitively, the simulated data is distributed more closely to the real data if more kernels are used in the model. 

A more clear demonstration of the characteristics of data distribution shown in Fig. \ref{fig:time_series} is given in Fig. \ref{fig:distribution_exa}. The simulator using 10 Gaussian kernels exhibits better fitting performance. The quantitative results of the fitting performance evaluation are plotted in Fig. \ref{fig:error_steady}. The statistics of the fitting error for each simulator are plotted in Fig. \ref{fig:error_steady} a). Instead of using model-estimated output as $\hat{o}$, the cumulative probability calculated from the simulated data is used. Since each simulator generated 500 groups in each validation simulation, the mean and variance of the fitting error were calculated. The solid line with circles represents the mean error and the dashed lines is the bounds of 99$\%$ confidence interval. The resulted plots show that mixture models exhibit much better fitting performance than the single Gaussian model. Besides, Fig. \ref{fig:error_steady} b) gives the Kolmogorov-Smirnov statistics which is expressed as\cite{Stephens}
\begin{equation}
\label{eq:kss}
D=\sup_{x}\|o(x)-\hat{o}(x)\|.
\end{equation}
The results of the Kolmogorov-Smirnov statistics have the same trend as the fitting error. Fitting error represents the mean square error and the Kolmogorov-Smirnov statistics show the maximal absolute value of error. Increasing the kernel number can decrease both.

\subsection{Transient case}
\begin{figure}[!htbp]
\centering
\includegraphics[width=4.5in]{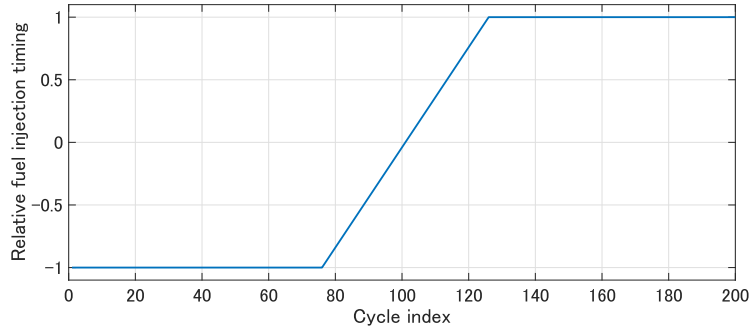}
\caption{Relative fuel injection timing evolution of the transient case.}
\label{fig:con_transient}
\end{figure}
\begin{figure}[!htbp]
\centering
\includegraphics[width=4.5in]{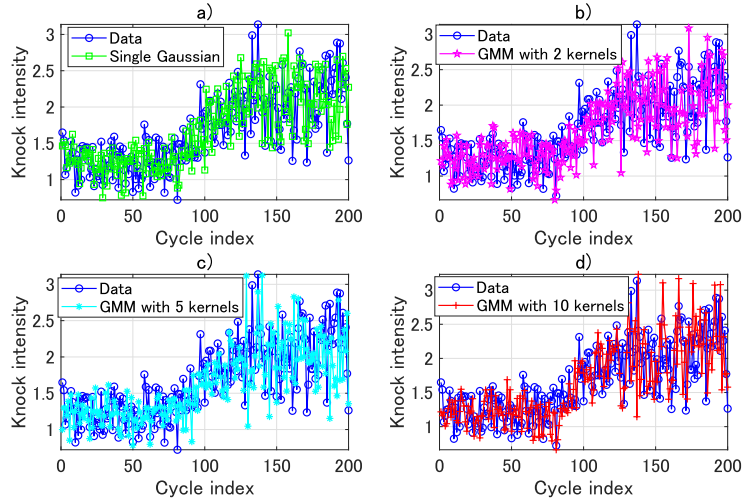}
\caption{Data of test set vs simulated data corresponding to the condition shown in Fig. \ref{fig:con_transient}: a) Single Gaussian model; b) GMM with 2 kernels; c) GMM with 5 kernels; d) GMM with 10 kernels.}
\label{fig:result_transient}
\end{figure}

In the transient case, the engine was operated at 1200 rpm. The manifold pressure is 7 bar. The relative fuel injection timing was firstly kept as constant and then changed to another constant. Fig. \ref{fig:con_transient} shows the relative fuel injection timing evolution of the transient case. The corresponding test set data and simulated data by several different simulators are shown in Fig. \ref{fig:result_transient}. Mixture models exhibit better performance than the single Gaussian model.

\section{Conclusion}
\label{section:conclusion}

This paper proposes a statistical knock simulator based on the mixture density network and the accept-reject algorithm. The knock intensity is a stochastic process. With a identical input, it is independent and identically distributed. Moreover, the distribution is a function of the input. The mixture density network is applied to approximate the function from input signal to the knock intensity distribution. The accept-reject algorithm is used to generate knock intensity according to the mixture density network-based knock intensity distribution model. The proposed method is evaluated in experimental data-based validation. Using mixture models can improve the performance of the simulator. With more kernels, the simulator is able to output the knock intensity that has closer distribution with the real data. However, there is still future work should be done. Especially, how to quantitatively evaluate the simulator performance for the transient case.  


\end{document}